\documentclass[sigconf]{acmart}

\AtBeginDocument{%
  \providecommand\BibTeX{{%
    \normalfont B\kern-0.5em{\scshape i\kern-0.25em b}\kern-0.8em\TeX}}}

\setcopyright{acmcopyright}
\copyrightyear{2020} 
\acmYear{2020} 
\setcopyright{acmcopyright}\acmConference[Chinese CHI 2020]{The eighth International Workshop of Chinese CHI}{April 26, 2020}{Honolulu, HI, USA}
\acmBooktitle{The eighth International Workshop of Chinese CHI (Chinese CHI 2020), April 26, 2020, Honolulu, HI, USA}
\acmPrice{15.00}
\acmDOI{10.1145/3403676.3403679}
\acmISBN{978-1-4503-8815-3/20/04}

\usepackage{wrapfig}

\begin{document}

\title{Friend Network as Gatekeeper: A Study of WeChat Users' Consumption of Friend-Curated Contents}


\author{Quan Li}
\affiliation{%
  \institution{AI Group, WeBank}
  \city{Shenzhen}
  \state{Guangdong}
  \country{China}
}\email{forrestli@webank.com}

\author{Zhenhui Peng}
\affiliation{%
  \institution{The Hong Kong University of Science and Technology}
  \city{Hong Kong}
  \country{China}
}\email{zpengab@connect.ust.hk}

\author{Haipeng Zeng}
\affiliation{%
  \institution{The Hong Kong University of Science and Technology}
  \city{Hong Kong}
  \country{China}
}\email{hzengac@connect.ust.hk}

\author{Qiaoan Chen}
\affiliation{%
  \institution{WeChat, Tencent}
  \city{Shenzhen}
  \state{Guangdong}
  \country{China}
}\email{kazechen@tencent.com}

\author{Lingling Yi}
\affiliation{%
  \institution{WeChat, Tencent}
  \city{Shenzhen}
  \state{Guangdong}
  \country{China}
}\email{chrisyi@tencent.com}

\author{Ziming Wu}
\affiliation{%
  \institution{The Hong Kong University of Science and Technology}
  \city{Hong Kong}
  \country{China}
}\email{zwual@connect.ust.hk}

\author{Xiaojuan Ma}
\affiliation{%
  \institution{The Hong Kong University of Science and Technology}
  \city{Hong Kong}
  \country{China}
}\email{mxj@cse.ust.hk}

\author{Tianjian Chen}
\affiliation{%
  \institution{AI Group, WeBank}
  \city{Shenzhen}
  \state{Guangdong}
  \country{China}
}\email{tobychen@webank.com}

\def\plainkeywords{Friend-curated content; information consumption; gatekeeper}
\renewcommand{\shortauthors}{Quan Li, Zhenhui Peng, Haipeng Zeng, Qiaoan Chen, Lingling Yi, Ziming Wu, Xiaojuan Ma and Tianjian Chen}

\begin{abstract}
Social media enables users to publish, disseminate, and access information easily. The downside is that it has fewer gatekeepers of what content is allowed to enter public circulation than the traditional media. In this paper, we present preliminary empirical findings from WeChat, a popular messaging app of the Chinese, indicating that social media users leverage their friend networks collectively as latent, dynamic gatekeepers for content consumption. Taking a mixed-methods approach, we analyze over seven million users' information consumption behaviors on WeChat and conduct an online survey of $216$ users. Both quantitative and qualitative evidence suggests that friend network indeed acts as a gatekeeper in social media. Shifting from what should be produced that gatekeepers used to decide, friend network helps separate the worthy from the unworthy for individual information consumption, and its structure and dynamics that play an important role in gatekeeping may inspire the future design of socio-technical systems.
\end{abstract}




\maketitle

\section{Introduction}
\par Social media services such as Twitter, Facebook, and WeChat\footnote{https://www.wechat.com/en/} empower millions of users to consume content from and disseminate information to their social counterparts. For example, WeChat, one of the most popular friend-based social media services in China, generates and circulates over $1.5$ million articles in the form of embedded posts or external links~\cite{knobloch2005impact,li2018weseer}. Given the abundant content on social media, users face challenges of identifying authentic and high-quality information~\cite{potter2007media}. Take WeChat as an example, some public accounts on it mainly publish content that can be easily monetized to grab the audience's eyeballs but lacks substance~\cite{zeng2017social}. 

\par One way to safeguard the integrity of information for users is developing algorithms to remove fake news and promote high-quality ones~\cite{eslami2015always,rader2015understanding}. However, these algorithms could not present precisely what a specific user is interested in the current stage. As an another means, many social media services offer the ``share'' features for users to spread information in their social circles. These users are called ``gatekeepers'', who pass along and comment on already available news items based on their interests \cite{twitter2012, gatekeeper2008}. Previous research on social media has shown how the gatekeepers on Twitter affect the audiences' information selection during a special event (e.g., 2009 Israel-Gaza conflict \cite{twitter2012}), and how users play roles of the gatekeepers to control information flows on Reddit~\cite{leavitt2017role}. However, unlike Twitter and Reddit in which gatekeepers often have no close relationship with users (e.g., a famous star as the gatekeeper), WeChat builds friend-based social media in which a user ideally knows all members in his/her circle. Bakshy et al. demonstrated that friends can expose individuals to cross-cutting content on Facebook \cite{facebook2015}. Nevertheless, WeChat does not have algorithmically ranked News Feed as Facebook does, but present contents shared by friends in an ordered timeline manner. Moreover, most of the WeChat users are from China, and they have a different cultural background than the users of Facebook who mainly come from western countries. There is a lack of understanding of how the friend network acting as a gatekeeper on WeChat affects the users' content curation behaviors. Such an understanding is important as it can not only help the content creators to learn how their works are spread in the friend-based social network but also facilitate such social media platforms to manage the information flow.

\par In this paper, we use WeChat as a lens to investigate how users leverage their friend networks as latent gatekeepers for content curations on the friend-based social media. Specifically, we reveal how WeChat users exploit the composition and tie strength of their friend network to safeguard the relevance, importance, popularity, and/or quality of information they consume. We further examine how users adopt the gatekeeping mechanism according to the changes of friend networks and interests over an extended period of time. We take a mixed-methods approach~\cite{wisdom2013mixed} to study the possible ``friend network as a latent gatekeeper'' phenomenon on WeChat. On the one hand, we quantitatively analyze over seven million WeChat users' reading behaviors and infer how these users accommodate and safeguard their varying information needs through different friend communities and social ties. On the other hand, we conduct an online survey with $216$ participants to qualitatively understand how and why users view the gatekeepers in their networks. In general, we find that users tend to turn to the friend network for information consumption if there is overloading information. They tend to exploit weak-ties getting exposed to new domains and turn to strong-ties when demanding credible and reliable information. Elder users with shorter WeChat experiences and fewer friends depend primarily on their friend network for information consumption. Users leverage social circles to gatekeep information interests and the interests and attention paid to them curated by one social circle can shift to another circle. The major contributions of this paper are as follows:
\begin{itemize}
\item We qualitatively and quantitatively study the potential phenomenon of WeChat users leveraging their friend network as a collective and dynamic latent gatekeeper.
\item We discuss the insights derived from our approach to inspire the future design of socio-technical systems.
\end{itemize}

\section{Background and Related Work}

\subsection{Gatekeeping in Social Media Era}
\par Unlike traditional media, today's social media can be indelibly remarked as We Media~\cite{bowman2003we}, i.e., user-operated media, in which there is no clear boundary between information producers, disseminators, and consumers, and the contents published are no longer constrained by length, timeliness, and the relevance to readers in a geographical and cultural sense~\cite{althaus2000patterns,galtung1965structure,mccombs1977predicting}. Although social media users play an active role in shaping the online information landscape~\cite{Shapiro1999Loneliness}, they might not have the same level of professional qualities as the experts in the media industry for ``gatekeeping'', i.e., scrutinizing content, safeguard its validity, veracity, and integrity before reaching the public~\cite{lewin1943forces}. Ira Basen~\cite{basen2011news} pointed out that digital media platforms have fewer filters and gates than traditional media, making it challenging for users to determine what is new and what is important. Keen~\cite{keen2011cult} mentioned that Web 2.0 has a negative impact on gatekeeping because of the reduction in gates or official gatekeepers who are accountable and professional. He maintained that ``\textit{gatekeepers are a necessity due to the flood of information coming digitally.}'' Clark~\cite{Clark2015Gatekeeping} interviewed several news professionals and asked how social media plays roles in their daily professional lives, showing that the downsizing of newsrooms has made an impact on the traditional role of the editors as a gatekeeper. Besides, different from the definition of ``gatekeeper'' for traditional media that in a sense if something is ``gatekept'', it won't go public to anyone, many social media services offer the ``share'' features for users to spread information in their social circles. In other words, even if a user decides not to share the content, the content could still be seen by its friends through other friends (if they choose to share the content). The above studies mainly focus on studying the changes of gatekeeping from traditional media era to today's social media era, under the context that social media consumers have to face a flood of fake news and information. Leavitt et al.~\cite{leavitt2017role} looked at Reddit to understand how the design of Reddit's platform impacts the information visibility in response to ongoing events in the context of controlling information flows (through gatekeeping). Similar to the functions of gatekeeping in social media platforms, social media influencers (SMIs) represent a new type of independent third party endorser who shape audience attitudes through blogs, tweets, and the use of other social media~\cite{abidin2017familygoals,freberg2011social,khamis2017self,matias2017followbias}, and there are technologies developed to identify and track the influencers. However, different from social influencers, the gatekeeping in social platform plays a latent role in a collective and dynamic manner. In our work, we focus on Moments - a distinguishing feature of friend network in WeChat, and study how WeChat users utilize their friend networks as latent gatekeepers collectively and dynamically to safeguard the information they consume.

\subsection{Algorithmic Content Curation on Social Media}
\par People are increasingly relying on online socio-technical systems that employ algorithmic content curation to organize, select and present information. Several studies have addressed customers' perception of automated curation~\cite{eslami2015always,rader2015understanding}. For example, Rader et al.~\cite{rader2015understanding} investigated user understanding of algorithmic curation in Facebook's News Feed through an online survey. They found that over $60$\% of the respondents implicate that the algorithmic News Feed caused them to miss posts from friends yet they still believe the algorithm that prioritizes posts for display in the News Feed. Similarly, Eslami et al.~\cite{eslami2015always} conducted a user study with $40$ Facebook users to examine their perceptions of the Facebook news feed curation algorithm. In contrast with the above algorithmic curation, in this paper, we study social media users' consumption of contents that come directly from their friend networks in the absence of any automated curation. Unlike algorithmic curation that arranges and ranks the news items based on designated features, consumption of friend-curated content on WeChat offers users complete controls over information selection. It would be interesting to have an in-depth analysis of the users' internal ranking scheme of friend-curated content.

\subsection{Factors that Affect Users' Content Curation}
\par Social media is one essential way for people to curate information and previous literature has studied several factors that affect users' content curation behaviors. For example, Leskovec et al.~\cite{leskovec2009meme} studied the spread of news across websites and found that blogs generally lag only a few hours behind mainstream news sites. Agrawal et al.~\cite{agarwal2008identifying} proposed a model to identify influential blog contributors. They found that the number of times a blog post is shared and the number of comments on it generates are positively related to the influence of its contributors. Khan~\cite{khan2017social} conducted an online survey that covered 1143 registered YouTube users and identified the factors that motivate user participation and consumption on YouTube through regression analysis. 
User-user relationship with various strengths is also one important factor that influence user's information seeking experience~\cite{arnaboldi2013egocentric,berkovsky2012personalized,gilbert2012predicting,gilbert2009predicting,kahanda2009using,panovich2012tie,petroczi2007measuring,wu2010detecting,xiang2010modeling,zhuang2011modeling}.
Gilbert et al.~\cite{gilbert2009predicting} bridged the gap between social theory and social practice by predicting the strength of interpersonal relationships in social media and conducting user study-based experiments on over 2000 social media relationships. Wu et al.~\cite{wu2010detecting} identified the two different types of intimate relationships among employees in enterprise social networks. Granovetter, M.S.~\cite{granovetter1977strength} proposed ``weak-ties'', which he believed can break through the social circles formed by strong-ties, enabling us to reach a diverse group of people and information. On the contrary, Krackhardt, D.~\cite{krackhardt2003strength} believed that strong-ties are the bonds of trust between people, so they are more willing to accept information brought by strong-ties than weak ones. In addition to qualitative research, many scholars leveraged data models as tools to quantify the relationship between social influence and the scale of information propagation~\cite{leskovec2007dynamics,sun2009gesundheit,wei2010diffusion}. 
However, similar user-user relationship research on WeChat is still relatively scarce. As one complement to the works above, 
we leverage a mixed-methods approach empirically to exploring how WeChat users exploit the composition and tie strength of their friend network to safeguard the relevance, importance, popularity, and/or quality of information they consume. It would be interesting to see whether and how different social tie strength can be reflected as gatekeepers in content curation.

\section{Methods}
\subsection{Article Reading and Gatekeeping on WeChat}
\par
\begin{figure}[h]
	\includegraphics[width=\linewidth]{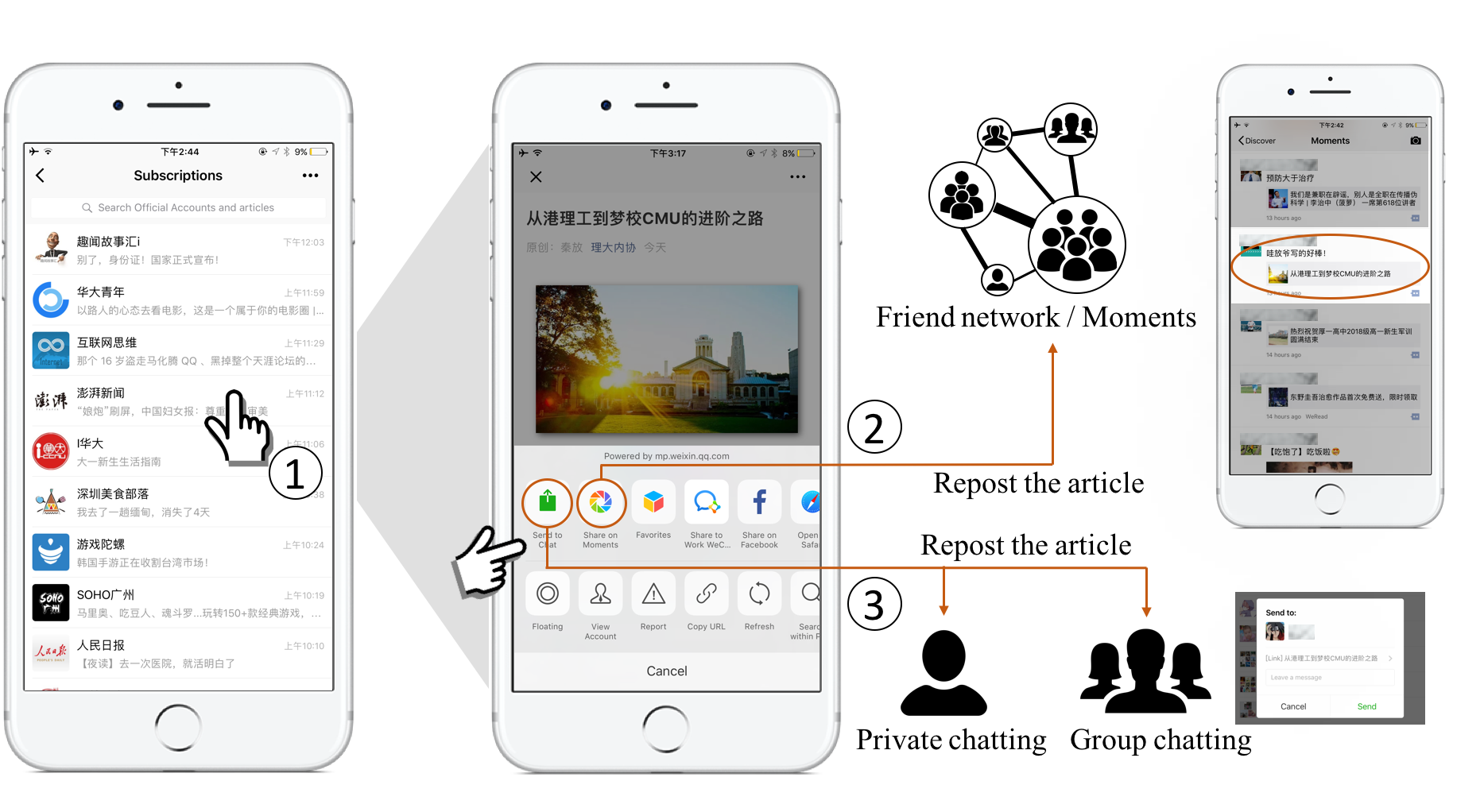}
	\caption{Article reading on WeChat. (1) Subscription accounts publish articles like news feed. (2) Users can repost articles on their Moments. (3) Users can repost articles via private chatting or group chatting.}
	\label{fig:one}
\end{figure}

 As a prevailing social media App, WeChat has its distinctive features. First, it owns the characteristics of traditional media. Users can read articles directly from the homepage of the subscription accounts from the WeChat Official Account Platform (\autoref{fig:one}(1)), which serves as the main source of articles for publication. Subscription accounts are often used similarly to daily news feeds because they can push one or several new update(s) to their followers every day~\cite{li2018weseer}. The update(s) could contain a single article or multiple articles bundled together. Users may subscribe to as many accounts as they like. All subscription accounts are placed together in a subscription accounts folder on the timeline of users. Similar to bloggers on Twitter, a WeChat subscription account also has its fixed author(s). Second, it owns the typical features of social media. The most intuitive feature is that users can read articles shared by and forward them to their friend network (\autoref{fig:one}(2)). WeChat provides three channels, namely, Moments, private chatting, and group chatting for users to access and read articles curated by their friends. In this work, we focus on content curation through the Moments. Users can share articles through their Moments, an immensely popular feature used to share pictures, short videos, texts, and links with their friends. Users can scroll through this stream of contents, similarly to Facebook newsfeed but they appear in chronological order. Users can also share articles to a specified friend or a group of friends via direct private or group chatting (\autoref{fig:one}(3)).

\par
\begin{figure*}[h]
	\includegraphics[width=\linewidth]{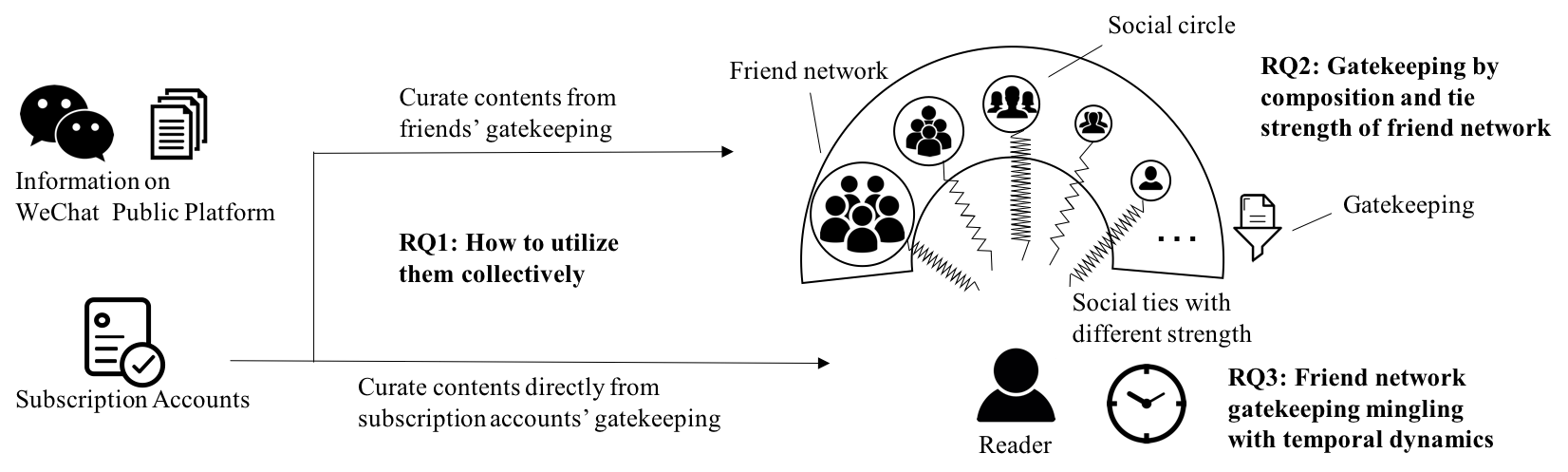}
	\caption{Gatekeeping process on WeChat and research questions in our study.}
	\label{fig:workflow}
\end{figure*}

 The gatekeeping process on WeChat is depicted in Figure~\ref{fig:workflow}. Given the large amount of information from the Internet, the hosts of subscription accounts act as the first level of gatekeepers. Users who want to curate content can directly read articles from these accounts, or they can read articles shared by their friends in the Moments. In the latter case, the user's friend network is acting as the second level of gatekeepers. The  subscription accounts and the friend network determine collectively which articles in the whole platform get to appear on individual users' news feed. 

\subsection{Research Questions}
\par To understand how users leverage their friend networks as latent gatekeepers for content curations, we first need to examine to what extent and in what way the subscription accounts and friend networks act collectively as gatekeepers for users. 
Therefore we have our first research question:

\begin{wrapfigure}{l}{0.03\textwidth}
  \vspace{-10pt}
  \begin{center}
    \includegraphics[width=0.08\textwidth]{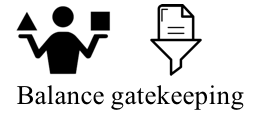}
  \end{center}
  \vspace{0pt}
  \vspace{-10pt}
\end{wrapfigure}

\textbf{RQ1. How do WeChat users utilize friend networks and subscriptions collectively as latent gatekeepers for content consumption?}

\par Li et al. revealed that WeChat users are often confronted by abundant friend-curated content from a wide variety of sources~\cite{li2018weseer}.  
Users may need additional cues to reduce the cognitive burden of deciding what to read. One potentially useful and always available cue is the composition (e.g., classmate, relative) and tie strength (e.g., how close is the relationship) of their friend networks. If a close friend is believed to be highly knowledgeable or trustworthy about public affairs, these positive evaluations may transfer to the information curated by him/her. We have our second question regarding the composition and tie strength of the friend network:

\begin{wrapfigure}{l}{0.03\textwidth}
  \vspace{-10pt}
  \begin{center}
    \includegraphics[width=0.08\textwidth]{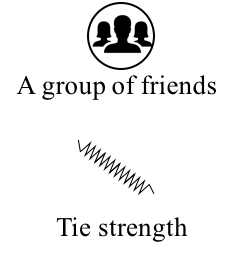}
  \end{center}
  \vspace{-10pt}
\end{wrapfigure}

\textbf{RQ2. Any difference between \textbf{a) }social circles and\textbf{ b)} social ties when acting as gatekeepers?}
\par Noted that the social contacts and information interests can change noticeably over time. Understanding these dynamics allows us to leverage those facets to improve relevance, and better manage influence and different ``gatekeepers'' in information dissemination~\cite{wang2016measuring}. Therefore, we study the third research question regarding temporal dynamics in the gatekeeping process:
\begin{wrapfigure}{l}{0.03\textwidth}
  \vspace{-10pt}
  \begin{center}
    \includegraphics[width=0.08\textwidth]{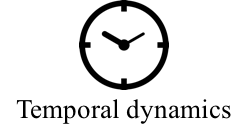}
  \end{center}
  \vspace{0pt}
  \vspace{-15pt}
\end{wrapfigure}

\par \textbf{RQ3. How do WeChat Users adapt gatekeeping for content consumption over time?}

\subsection{Quantitative Analysis Method}
\par Archival data obtained through collaboration with WeChat reveal that messages coming from all WeChat channels, i.e., subscription accounts, private chatting, group chatting, and friend network (i.e., Moments), collectively create users' information landscape on the platform, each taking up different proportions. On average, 57\% of the articles consumed by a WeChat user come directly from subscription accounts. The remaining 43\% are shared by friends through private chatting (11\%), group chatting (18\%), and Moments (71\%) under different social scenes~\cite{zhang2018mobile}. Private chatting ``digests'' the articles exchanged in the communication~\cite{wu2014wechat}. Group chatting is a private conversation among a group of users pre-gathered for certain purposes. Note that members of a WeChat group may not necessarily be friends of one another. Therefore, it is a social environment with complicated and unpredictable factors~\cite{qiu2016lifecycle}. Comparatively, Moments is like a public bulletin for one's entire friend network, publishing all the contents posted by friends in a timeline manner. Theoretically, people can browse content on their Moments at will, similar to how they can treat posts curated in the subscription account folder (if we consider subscription accounts as a special type of ``friend''). As the focus of this paper is on how users proactively leverage their friend network as a latent gatekeeper of their information landscape, we only consider voluntary reading behaviors related to the two broadcasting channels, i.e., subscription accounts and Moments.

\subsubsection{Data collection and Description}
\par The dataset in this work was collected by our collaboration colleagues from WeChat, Tencent. Particularly, the dataset used for \textbf{RQ1} and \textbf{RQ2} contains a one-week log of article-reading activities from March $12^{th}$ - $18^{th}$, 2018 curated via the subscription accounts and friend network of 7,234,753 users, a stochastic sampling on all users. The dataset is anonymized with all identifiable information removed. It consists of three parts: 
\begin{itemize}
	\item \textbf{A1. User Attributes} include user information within the selected time frame, such as age, registration duration, the number of friend, and the list of official accounts subscribed. 
	\item \textbf{A2. User Social Relationship} contains the list of friends of each user and their social relationship with the users. In this paper, we describe a social relationship through the following two dimensions: 
	\begin{itemize}
		\item \textbf{\textit{D1. Social Similarity}} calculates the number of common friend between two users, which is a common practice to indicate to what extent two users are similar in social network analysis~\cite{banks2008social}. We thereby adopt it to obtain the social similarity between two users.
		\item \textbf{\textit{D2. Social Circle}} presents social community in this work. Our collaboration experts generate four types of labels for communities in one's friend network: colleagues, family, schoolmates, and others (e.g., real estate agency and WeChat business) by adopting a community detection algorithm \textit{Fast Unfolding}~\cite{blondel2008fast}.
	\end{itemize}
	\item \textbf{A3. User Article Consumption} includes (1) the list of articles published by all the subscription accounts that a user follows; (2) the list of articles consumed by the user from the subscription accounts; (3) the list of articles curated by the friend network; and (4) the list of articles consumed by the user from the friend network. Note that these data can be filtered based on the attributes D1 and D2.
\end{itemize}
\par To answer \textbf{RQ3}, we collect an additional one-year article-reading data (201709 - 201807) of $10,000$ users via stochastic sampling on WeChat user pool. Compared with the previous one-week dataset for \textbf{RQ1} and \textbf{RQ2}, this one-year sampling dataset only contains \textit{A1. User Attributes} and \textit{D2. Social Circles}. This dataset has an additional attribute that is not included in the one-week dataset: 
\begin{itemize}
	\item \textbf{A4. Article Categories} specify the aggregated number of articles each user consumes in terms of different article categories.
\end{itemize}

\subsubsection{Preliminaries and Computational Metrics}
\par In this subsection, we provide a brief overview of the definition of metrics and terms used in the quantitative analysis for \textbf{RQ1} and \textbf{RQ2} (note: metrics for \textbf{RQ3} are described in \textbf{RQ3} subsection in \textit{RESULT} section). We define information consumption on WeChat as a behavior of user clicking on a certain article curated by friends that shows up in a user's Moments. To the user, these friends are gatekeeper of his/her Moments, determining what can be circulated in it. We define:
\begin{itemize}
	\item \textbf{M1. Click-through Rate (CTR)} is the ratio of the number of consumed articles to the number of exposed articles for a WeChat user.
	\item \textbf{M2. Influence Ratio (IR)} is the influence ratio of user $i$ to user $j$ which is measured as:
	\begin{equation}
	r_{i->j} = \frac{m_{i->j}}{n_i}
	\end{equation}
	where $m_{i->j}$ is the number of times user $j$ read articles shared by user $i$ and $n_i$ is the total number of articles that user $i$ share. The IR quantifies the pairwise influence between the user and his/her friend. The larger the IR, the more attention user $j$ pays to contents curated by user $i$, and thus the greater influence user $i$ has on user $j$. 
	\item \textbf{M3. Total influence of friends' gatekeeping} indicates the total effect of friends' gatekeeping on user $j$ which is defined as: 
	\begin{equation}
	r_j = \frac{\sum_{i \in F_j}{m_{i->j}}}{\sum_{i \in F_j}n_i}
	\end{equation}
	where $F_j$ is the set of friends of users $j$.
	\item \textbf{M4. Influence of a social circle} indicates the influence of a certain type of social circle $E$ which is defined as: 
	\begin{equation}
	r_E = \frac{m_E}{n_E} = \frac{\sum_{(i,j)\in E}(m_{i->j}+m_{j->i})}{\sum_{i \in V(E)}n_i}
	\end{equation} where $V(E)$ is the involved users in the set of friends $E$.
	\item \textbf{M5. Influence of subscription accounts on user} $j$ is defined as: 
	\begin{equation}
	s_j = \frac{k_j}{l_j}
	\end{equation}
	where $l_j$ is the total number of articles published by all the subscription accounts that user $j$ follow, and $k_j$ is the total number of articles that user $j$ reads directly from these subscription accounts. 
	\item \textbf{M6. The ratio of the influence of subscription $s$ over the influence of friends $r$}: $\frac{s}{r}$ describes how users split the gatekeeping responsibilities between subscription accounts and the friend network.
\end{itemize}

\subsection{Qualitative Analysis Method}
\par To verify the potential quantitative results with the users and to understand why users have some specific behaviors for gatekeeping of friend networks, we conduct an additional qualitative analysis with WeChat users. 

\subsubsection{Participants}
\par We recruited $216$ participants (males: 53.7\%, females: 46.3\%; age (19-25): 24.0\%, age (26-30): 28.2\%, age (31-40): 24.5\%, age (41-60): 20.8\% and age (60+): 2.50\%) via an online survey service to understand their information consumption experience on WeChat. All the survey participants have a good knowledge of WeChat, as well as information consumption on WeChat. Particularly, we choose the participants with good operation skills of WeChat, for which they could provide us more comprehensive insights. We further invited $10$ survey respondents (P1-P10) (males: 60\%; females: 40\%; age (19-25): 20\%, age (26-30): 20\%, age (31-40): 30\%, age (41-60): 20\% and age (60+): 10\%) for follow-up interviews about their choices in the survey. Each interview took $15$ minutes and was audio recorded.

\subsubsection{Design of Questionnaires}
\par The questions used in the questionnaires take the form of multiple choices. As a supplement for \textbf{RQ1}, we ask participants to choose what kinds of articles are most welcome from their friend network and what kinds of articles they would further curate and repost. For \textbf{RQ2}, they are asked to indicate from which social circle(s) (options: family, colleague, schoolmate) do they curate information in their friend networks and choose the possible reasons we provide. For \textbf{RQ3}, we ask them whether and when would their social circles as ``gatekeepers'' experienced some changes.

\section{Results}
\par In this section, we summarize the results for the research questions one by one, following the style of first presenting the quantitative results then listing the qualitative one if applicable.

\begin{figure}[h]
\includegraphics[width=\linewidth]{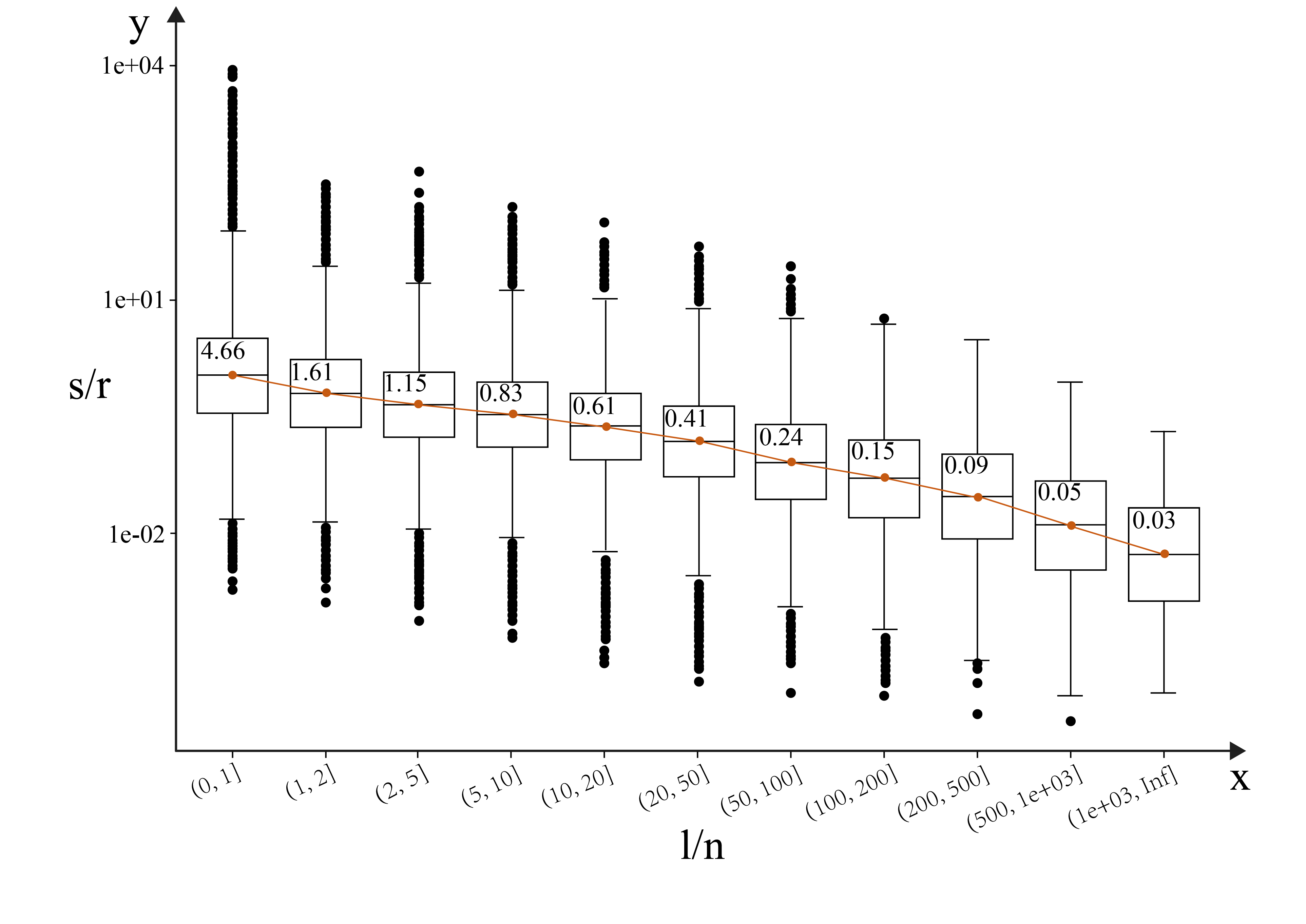}
\caption{X-axis: the ratio of the number of articles $l$ published by the subscription accounts over the number of articles $n$ curated by the friend network AND y-axis: the ratio of the influence of subscription accounts $s$ over the influence of the friend network $r$.}
\label{fig:rq11}
\end{figure}

\subsection{RQ1: Gatekeeping by Subscription Accounts and Friend Network Collectively}

\par \autoref{fig:rq11} shows the relationship between $\frac{s}{r}$ (the ratio of the influence of subscription $s$ over the influence of friends $r$) and $\frac{l}{n}$ (the ratio of the number of articles $l$ published by the subscription over the number of articles $n$ shared by friends). 
We can see that $\frac{s}{r}$ decreases with the increase in $\frac{l}{n}$. Noted that in \autoref{fig:rq11}, we adopt a logarithmic axis. One can see that the ratio of the influence of subscription accounts over the friend network has a power-law dependence on the ratio of the articles published by the subscription accounts over the ones by the friend network:

\begin{equation}
\frac{s}{r} \approx \beta(\frac{l}{n})^{\alpha}
\label{equ:5}
\end{equation}

\begin{figure*}[h]
\includegraphics[width=\linewidth]{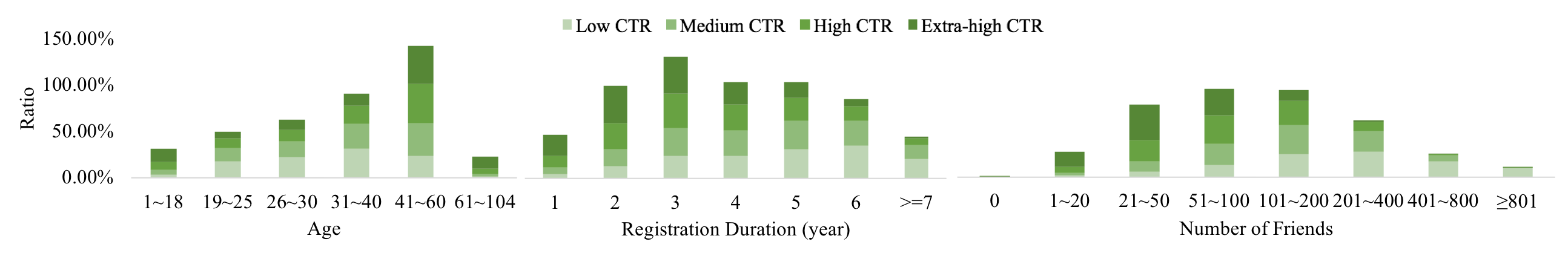}
\caption{Distribution of different age, registration duration, and number of friends over the four levels of click-through rate (CTR).}
\label{fig:rq13}
\end{figure*}

\par Through linear regression, we obtain $\alpha \approx -0.48$ and $\beta \approx 0.75$ with the significant level exceeding $99$\%. With a further deduction from Equation~\ref{equ:5}, we can infer that when $\frac{l}{n} > 0.55$, $\frac{s}{r} < 1$. That is: when the number of articles published by the subscription accounts exceeds 55\% of the number of articles shared by friends, the influence of the friend network will be greater than that of the subscription accounts (\textbf{Finding 1 (\textbf{F1})}).

\par Due to the zero-sum nature of attention~\cite{zhu1992issue}, WeChat users rely primarily on the subscription accounts and the friend network to filter information within their reach. \textbf{F1} suggests that users are likely to adjust their degree of reliance on each channel based on the quantity and quality of its content supply. If a user only subscribes to a few official accounts selectively, articles received from this channel are limited in quantity and more likely to catch the user's attention upon arrival. On the contrary, when articles from the subscription accounts are flooding the user's wall, the subscription channel can no longer help separate the attention-worthy content from the unworthy effectively. The user may instead turn to the friend network with finer ``gates'' to control the information flow.

\par To conduct an in-depth analysis of the target users who are likely to consume content from the friend network, we divide the value of \textbf{M1} CTR into four ranges: $0$ - $0.05$ as low CTR, $0.05$ - $0.15$ as medium CTR, $0.15$ - $0.3$ as high CTR, and over $0.3$ as extra-high CTR according to the input of domain experts from our collaborator. We then group users by their CTR range and plot the distribution of three user attributes \textbf{A1} in each group: \textit{age} ($1$ - $18$, $19$ - $25$, $26$ - $30$, $31$ - $40$, $41$ - $60$ and over $61$), \textit{registration duration} (years), and \textit{the number of friends} with the ranges of $0$, $1$ - $20$, $21$ - $50$, $51$ - $100$, $101$ - $200$, $201$ - $400$, and $401$ - $800$. \autoref{fig:rq13} (left) shows the distribution of different age groups over the four CTR groups. The $41$-$60$ age group has the highest percentage in each CTR group, especially in high ($42.57$\%) and extra-high CTR ($41.44$\%), largely surpassing the other CTR groups in this age range. The distribution of registration period over the four CTR categories (\autoref{fig:rq13} (middle)) shows that the higher CTR user groups tend to have a shorter registration history. We also find that users of high or extra-high CTR tend to have fewer friends on WeChat (noted as \textbf{F2}) (\autoref{fig:rq13} (right)).

\par The qualitative results for \textbf{RQ1} show the diversity of the curated and gatekeeping content. 79.6\% of the survey respondents stated that they often consumed articles on WeChat. Articles with \textit{attractive titles} (42.6\%), \textit{news and events} (38\%), \textit{practical knowledge} (34.7\%), \textit{financial and investment knowledge} (26.4\%), and \textit{funny stories} (25.5\%) are most welcome from the friend network. Users would further curate and repost the articles about \textit{practical knowledge} (56\%), \textit{insightful stories} (28.7\%), \textit{industry trends} (27.8\%), \textit{``chicken-soup'' articles} (nourishing stories for one's soul) (26.4\%), and \textit{current events} (21.8\%), etc., a bit different from what they consume from the friend network. Respondents (P7, P9, females; P3, P5, males) stated that ``\textit{sometimes, these articles represent what we thought,}'' and ``\textit{are well responsive to our current status.}'' Responses from different participants indicate different media literacy, showing that contents curated by different friends can be quite diverse.

\subsection{RQ2: Gatekeeping by Composition and Tie Strength of Friend Network}

\begin{figure}[h]
\includegraphics[width=\linewidth]{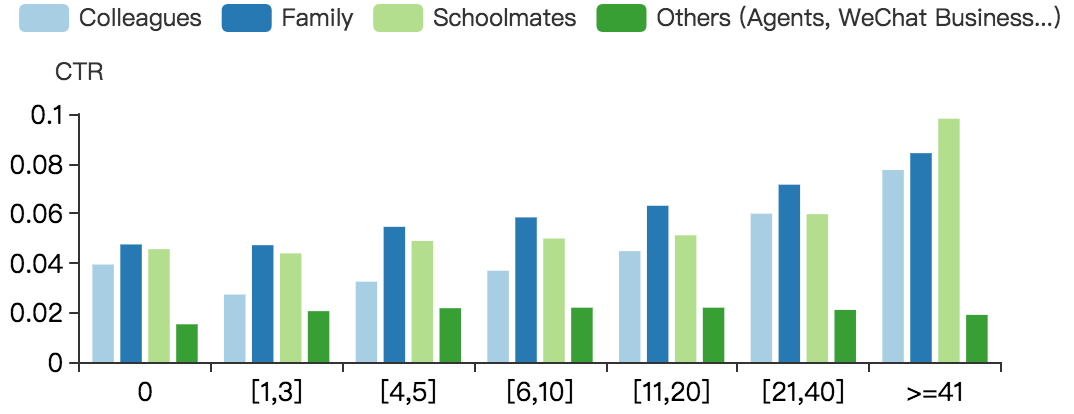}
\caption{In overall, with the increase of x-axis: social similarity, y-axis: CTR also increases.}
\label{fig:rq21}
\end{figure}

\subsubsection{RQ2a: About Social Circle}
\par
\begin{figure}[h]
\includegraphics[width=\linewidth]{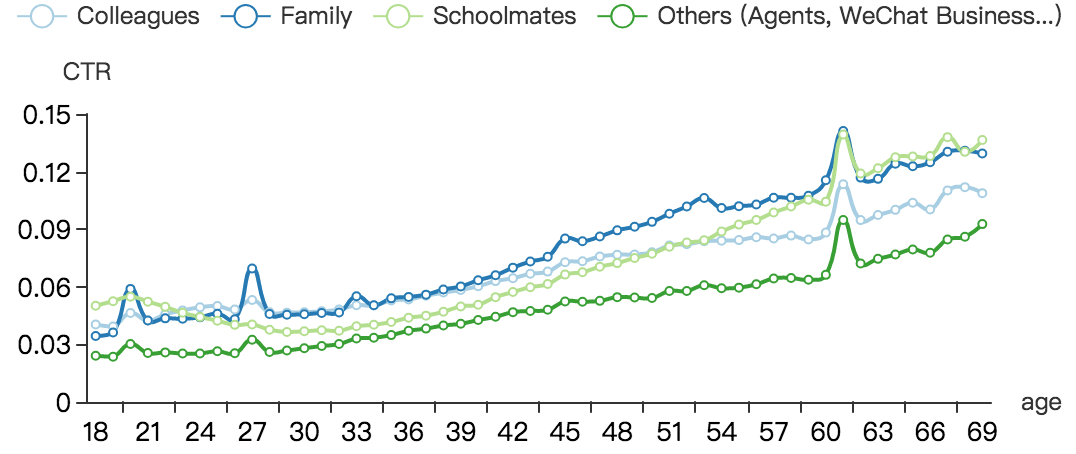}
\caption{CTR of four social circles (i.e., colleagues, family, schoolmates, and others over the ages from 18 to 69).}
\label{fig:rq24}
\end{figure}

 To explore how the composition of friend network help WeChat users filter information in Moments, we analyze to what extent friends with different social attributes (\textit{D1. Social Similarity} and \textit{D2. Social Circle}) serve as users' channel of choice for content consumption (measured by CTR). We divide the value range of social similarity (number of common friends) into seven intervals: $0$, $1$ - $3$, $4$ - $5$, $6$ - $10$, $11$ - $20$, $21$ - $40$, and over $40$ according to the input of domain experts from our collaborator. As shown in \autoref{fig:rq21}, CTR goes up as social similarity increases. Among individuals having over 40 common friends with a user, those from the user's schoolmates circle achieve the highest CTR. In the rest of the social similarity intervals, CTR of the family circle is the highest (noted as \textbf{F3}). When rendering the CTR values of friends from various social circles for users at different ages (\autoref{fig:rq24}), we find that users across all age groups generally prefer to consume articles curated by the family circle, followed by colleagues and schoolmates. Interestingly, the influence of schoolmates first declines among users at age 20 and then rises again among users aged 32 and over, eventually surpassing the CTR of colleagues among users at age 54 and exceeding the family CTR for users aged 60 or older. Colleagues have about the same level of influence as (sometimes slightly higher than) family for users in their 20s and 30s, and gradually lose the leading position to family and then to schoolmates among users over 40. In a word, the proportion of information intake from different social circles varies across WeChat users in different age groups (noted as \textbf{F4}).

\par \textbf{F3-4} manifests that people always care about family-curated content. However, 60.7\% of the survey respondents do not consume information from the family circle, especially from the elder ones. Among them, 72.5\% reported that they are not interested in the topics curated by the elder family members, and 29\% indicated that their reading interests are not similar. ``\textit{Family members like to share articles about inspiring stories, health maintenance, and festival-greetings, etc.,}'' said P3 (male). 29.4\% of those who consume family-curated articles ``\textit{because of emotional support}'' and 74.1\% of them ``\textit{care about what my family members are interested in.}'' Only 22.4\% reported that they share similar interests with their family. ``\textit{We may just take a glance at the title or quickly go through the contents,}'' said P7 and P9 (females). In this case, it seems that people may already have a pre-assumption for the information curated by their families. Consuming information from them may be largely due to emotional support, rather than the information relevance or importance. To further verify this hypothesis, we need more data such as the average time of reading an article curated by different circles.
Apart from the family circle, 74\% and 76\% of the survey participants like to consume information from colleagues and schoolmates because of the topics (71.5\%) and similar hobbies (46.7\%). 27.3\% of users reading articles from colleagues stated that ``\textit{these articles can be conversation-makers in the company.}'' 23.6\% of users (with 34.43\% from the age of $26$-$30$) who do not like to read articles from schoolmates stated that ``\textit{our life becomes different.}''

\subsubsection{RQ2b: About Social Tie Strength}
\par We then explore the use of tie strength of their friend network for filtering information in Moments. When Granovetter, M.S. first proposed the concepts of strong-/weak-ties, he did not provide a strict definition but a qualitative description: strong-ties refer to frequent connections and close relationships, and weak-ties are accidental connections with seldom communications~\cite{granovetter1977strength}. In this study, we follow the approach proposed by Gupte et al.~\cite{gupte2012measuring} and employ Jaccard Index\footnote{https://en.wikipedia.org/wiki/Jaccard\_index} of \textit{D1: Social Similarity} to measure the tie strength quantitatively:
\begin{equation}
J(F_i, F_j) = \frac{\lvert F_i \cap F_j \rvert}{\lvert F_i \cup F_j \rvert}
\end{equation} where $F_i$ is user $i$'s friends and $F_j$ is user $j$'s friends.

\begin{figure}[h]
\includegraphics[width=\linewidth]{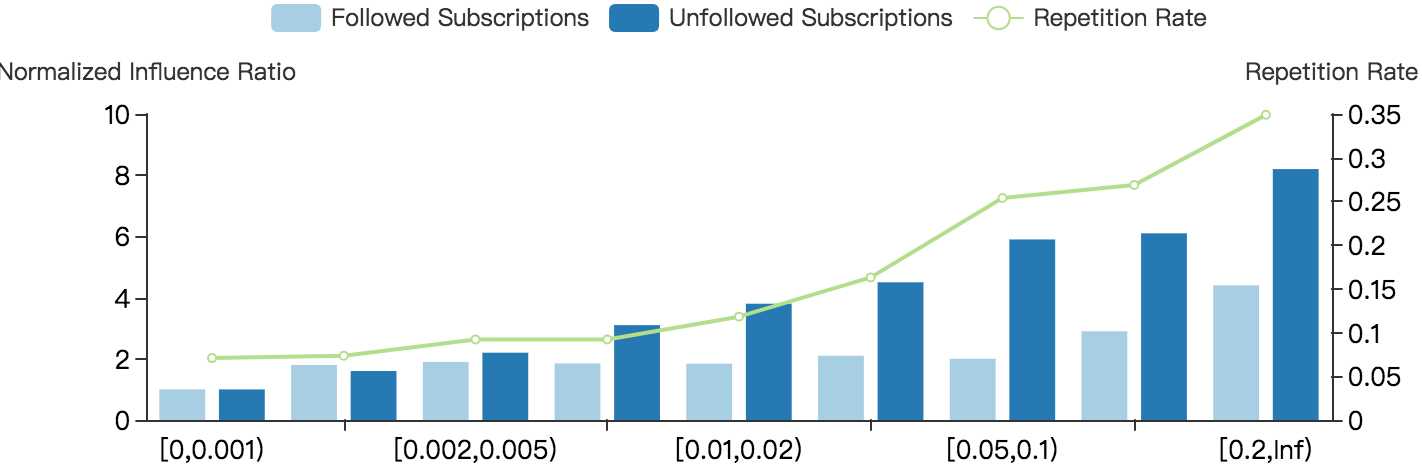}
\caption{Bars show relationships between tie strength (x-axis) and normalized influence ratio (left y-axis) from subscribed and unsubscribed cases. The curve shows relationships between tie strength and repetition rate (right y-axis).}
\label{fig:rq22}
\end{figure}

\begin{figure*}[h]
\includegraphics[width=\linewidth]{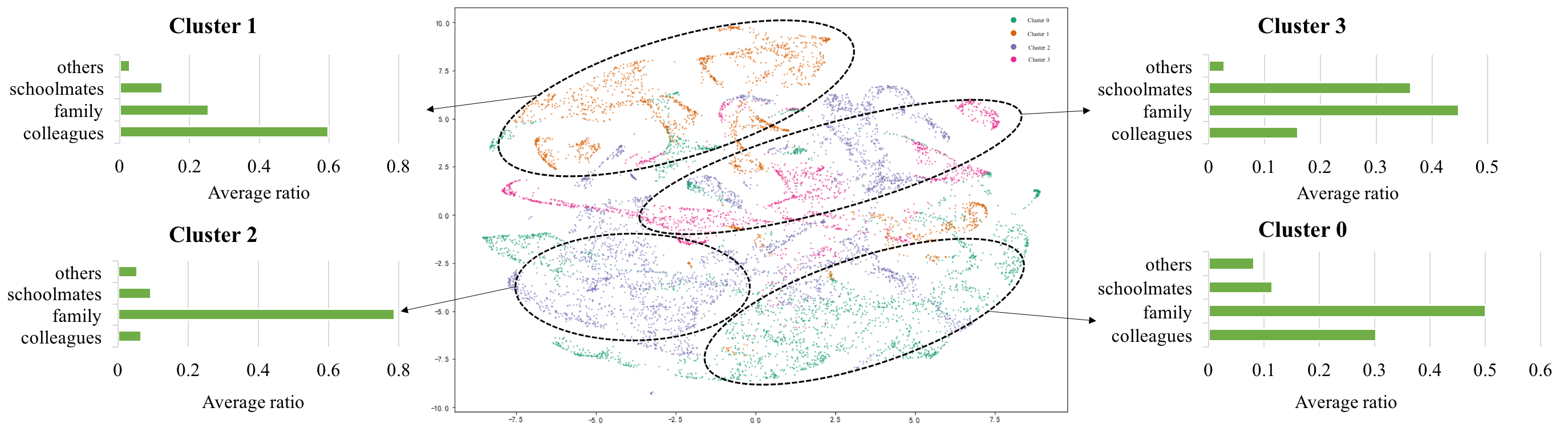}
\caption{T-SNE projection of all the sampled users. The average ratio is computed for each social circle in each cluster.}
\label{fig:rq31}
\end{figure*}

\par \textbf{Strong Ties Bring Trust.} Articles curated by friends on WeChat may be either from the subscription accounts that \textbf{(Case 1)} a user has already followed or from those \textbf{(Case 2)} the user has not yet followed. We utilize \textbf{A2-3} to compute and compare \textbf{M2. Influence Ratio} in the two cases, and apply normalization as follows. For each case, we divide the influence ratio in each segment of tie strength by the minimum influence ratio among all segments in that case, and then obtain the normalized influence ratios of the corresponding cases in each segment. As shown in \autoref{fig:rq22}, we find the influence of strong-ties among most segments in Case 2 is much higher than that in Case 1. For example, in Case 2, the influence ratio of the segment of [0.05, 0.1) is six times higher than that of the segment of [0, 0.001), whereas in Case 1, it is only two times. Case 2 confirms the ``strong-ties theory'' proposed by Krackhardt, D.: ``\textit{consuming articles from unknown sources means making changes (e.g., exposure to new knowledge, cognitive changes.), and it is with discomfort; however, strong-ties can help overcome this discomfort.}''~\cite{krackhardt2003strength} In Case 1, users have direct exposure to the articles from the subscription account. If they have determined to read the articles (or not), seeing the articles in their friends' Moments may not change their decision. This is perhaps why the normalized influence ratios for Case 1 are pretty similar between the $2^{nd}$ and $7^{th}$ bar. However, there is a noticeable increase of Case 1 influence ratios in the last two bars, suggesting the likely persuasion effect of the strongest ties. This finding indicates that strong-ties bring a sense of trust as a gatekeeper. In other words, if an article comes from a subscription account that a user has not followed, the trust brought by this article is deficient. However, the friend curation behavior makes up for this lack of trust and thus this behavior can be considered as a trustful gatekeeper (noted as \textbf{F5}).

\par \textbf{Weak Ties Bring Serendipity.} We investigate the repetition rate of the articles curated by the friend network in each segment of tie strength (we calculate the repetition rate which indicates the percentage of friends ever curating the same articles based on \textit{A2-3}) (the green curve in \autoref{fig:rq22}). We find that with the increase of tie strength, the repetition rate rises gradually. This indicates that weak-ties are more likely to bring information that users have not seen before, and can act as ``gates'' leading to ``unexpectedness'' (noted as \textbf{F6}). This finding also corroborates the ``weak-ties theory''~\cite{granovetter1977strength} that ``\textit{information curated by strong-ties is likely to be similar and redundant, whereas weak-ties can break the boundaries of people's inherent social circles and bring new information.}''

\par In the qualitative study, we asked the interviewees about how they treat the information curated by strong-/weak-ties. ``\textit{I want to learn why my close friends read this article,}'' said P6 (female), ``\textit{I will consume information from people I occasionally meet because of their comparatively fresh information.}'' ``\textit{I pay special attention to some friends who have special ideas, or opinion leaders,}'' said P9 (female).  ``\textit{I'm interested in articles from strong-ties, but sometimes, weak relationships will also bring some current affairs-related articles which I am interested in,}'' said P5 (male). ``\textit{Sometimes, I have a clear idea of what I want and go straight to appropriate friends for contents known to fulfill my consumption demand; other times, I am open to new information and just click around,}'' said P4 (male). When we further inquired him whether these friends could act as a ``filter'', he said, ``\textit{definitely, with so much information, I will choose the information curated by my close friends with trustworthiness.}''

\par 49.5\% survey respondents stated that they would follow up a popular event only after their friends have exploded with it, compared with that 32.4\% of the respondents would proactively seek relevant information in no time. From the interviews, we confirm that people may transfer positive evaluations from trustful friends to their curated information, ``\textit{I tend to believe what my friends believe,}'' said P7-8 (females), which is also consistent with our quantitative analysis of ``strong-ties bring trust'' (\textbf{F5}), i.e., an article curated by a close friend will increase the trust in the corresponding unfollowed subscription accounts. To further verify whether users will follow and consume information from these unfollowed subscription accounts, more data are needed.

\begin{figure*}[h]
\includegraphics[width=\linewidth]{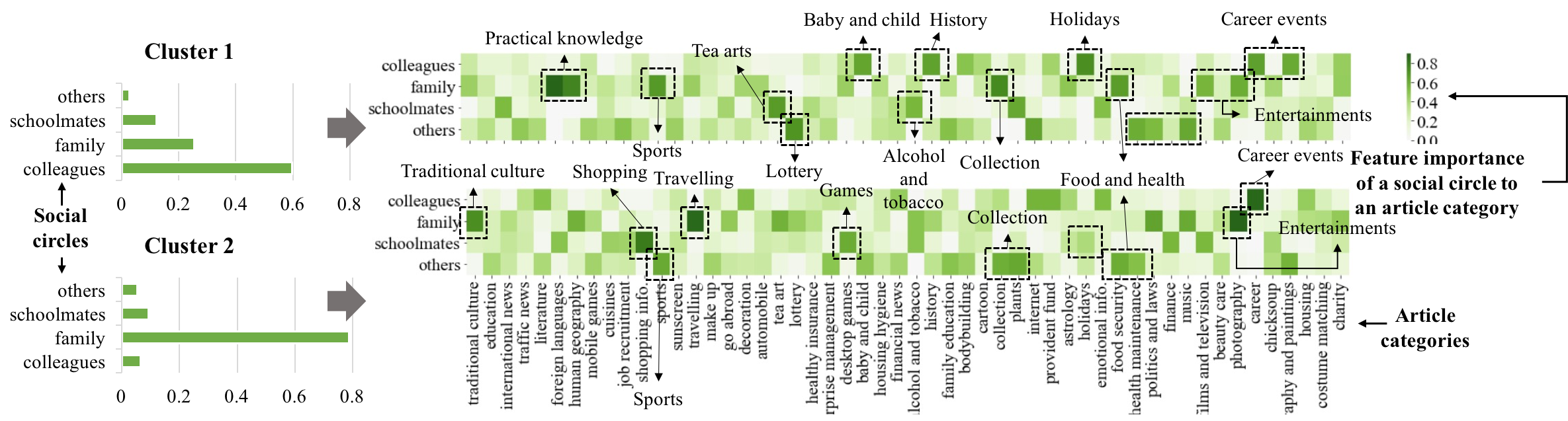}
\caption{Color blocks indicate feature importance of four social circles for different article categories. Cluster 1 (colleagues occupying the most) and Cluster 2 (family occupying the most) are compared using data of 201709.}
\label{fig:rq32}
\end{figure*}

\subsection{RQ3: Friend Network Gatekeeping Mingling with Temporal Dynamics}
\par We take two steps to address this research question.

\par \textbf{Step 1: Computing feature importance at each time frame.} We first derive a computational model to infer how WeChat users leverage different social circles to gatekeep the relevance/types of articles within their reach at each time frame of one-month. We start with depicting each user $u$ by its friend network composition (\textbf{D2}), $v_u = (r_{c}, r_{f}, r_{s}, r_{o})$, where $r_c$, $r_f$, $r_s$, $r_o$ represent the ratio of the number of friends in \textit{colleagues}, \textit{family}, \textit{schoolmates}, and \textit{others}, respectively, to all friends. Based on its vector representation $v_u$, we use K-Means to cluster all sampled users into four clusters. Each cluster indicates a different friend network composition. The number of clusters can be dynamically adjusted. As shown in \autoref{fig:rq31}, when k = 4, we can achieve a balanced distribution among all clusters. Next, via regression analysis, we fit the friend network composition space to the article consuming space. The regression analysis has long been used to model the relationship between variables and to estimate how a dependent variable responds to changes~\cite{li2018embeddingvis}. Under the assumption that network composition can affect article consumption behaviors, we regard each social circle as an observed variable and use its combination to regress the consumption space for a certain article category (\textbf{A4}). For each cluster, we weigh each social circle by its contribution to the article consumption by using feature importance, i.e., we approximate the two spaces by
$
f(n_k, n) \approx reg(w_{c}*r_{c}, w_{f}*r_{f}, w_{s}*p_{s}, w_{o}*p_{o})
$ where $f$ computes the ratio of the number of consumed articles ($n_k$) of category $k$ to all consumed articles ($n$) and $w$ is the feature importance.

\par In this step, we apply five widely-used machine learning algorithms, including \textit{Linear Regression (LR)}, \textit{Lasso}, \textit{Multiple-layer Perceptron (MLP)}, \textit{Decision Tree (DT)}, and \textit{Random Forest (RF)} to conduct regression analysis. Among them, LR and Lasso are linear regressors and the rest fit data with non-linear kernels. We use the coefficient of determination ($R^2$) commonly employed in regression analysis to assess model performance (Table~\ref{tab:result}). The results indicate that DT performs the best with a sufficiently high $R^2$ score.

\begin{table}[h]
\begin{tabular}{l|l|l|l|l}
{Regression} & {$R^2_{id=0}$} & {$R^2_{id=1}$} & {$R^2_{id=2}$} & {$R^2_{id=3}$} \\
\hline
{LR} & 0.102 & 0.145 & 0.079 & 0.089 \\
\hline
{Lasso} & 0.051 & 0.072 & 0.052 & 0.062 \\
\hline
{MLP} & 0.252 & 0.225 & 0.191 & 0.430 \\
\hline
\textbf{DT} & \textbf{0.761} & \textbf{0.742} & \textbf{0.543} & \textbf{0.801} \\
\hline
{RF} & {0.540} & {0.680} & {0.601} & {0.762}
\end{tabular}
\caption{Results for different regression models.}
\label{tab:result}
\end{table}

\begin{figure*}[h]
	\includegraphics[width=\linewidth]{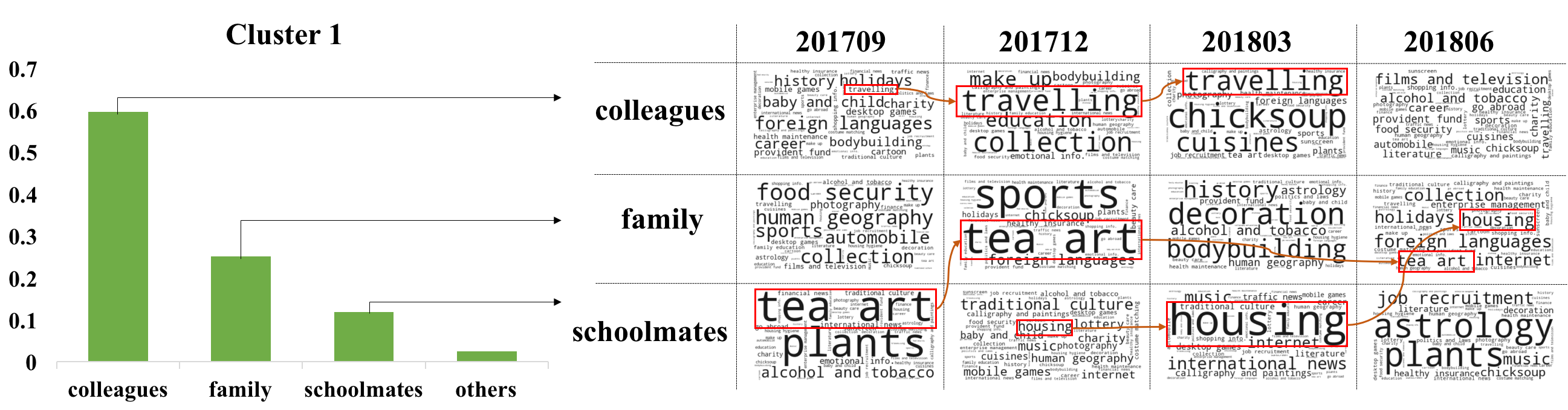}
	\caption{Word clouds indicate topics that social circles of Cluster 1 contribute to over time. The font size in word clouds encodes feature importance of the corresponding circle that contributes to the topic.}
	\label{fig:rq33}
\end{figure*}

\par We then apply DT to extract the feature importance of each social circle for each cluster to reflect their contribution to the circulation of each type of article. \autoref{fig:rq32} gives an example. In Cluster 1, colleagues occupy about 60\% of the friend network, followed by the family circle (25\%). Color blocks indicate the feature importance of the corresponding circles for different article categories. One can see that for Cluster 1, topics with high feature importance are \textit{practical knowledge} from family as well as \textit{baby and child}, \textit{holidays}, and \textit{career events} from colleagues. For Cluster 2 in which family circle dominates (80\%), the distribution of topics with high feature importance (e.g., \textit{traditional culture}, and \textit{traveling} from family, \textit{games} and \textit{shopping} from schoolmates) is different from that in Cluster 1. For a specific cluster of users, different social circle curates different topics (noted as \textbf{F7}). For example, \textit{alcohol and tobacco}-related articles come mostly from the schoolmate circle, and the schoolmate circle tends to circulate information about \textit{career events} and \textit{entertainment}. There is also the phenomenon that information about the same topic is curated by a different circle in different clusters. For example, in Cluster 1, the family is the main source of \textit{sports} and \textit{food}-related information, while such articles come mostly from ``others'' in Cluster 2. Another example is that users in Cluster 1 take in \textit{holiday}-related content primarily from colleagues, while those in Cluster 2 read about \textit{holidays} from their schoolmates' posts. This is mainly due to different clusters of users who may have quite different media literacy~\cite{potter2007media}.

\par \textbf{Step 2: Visualizing feature importance over time.} After calculating the feature importance for each month, we visualize the topics with feature importance encoded by font size in a word cloud. Take three social circles of users from Cluster 1 at four different time frames (i.e., 201709, 201712, 201803, and 201806) as an example (\autoref{fig:rq33}). Apart from some topics that are always dominated by certain social circles, e.g., \textit{traveling} for colleagues and \textit{plants} for schoolmates, \textit{housing} emerges on Dec. 2017 in the schoolmate circle and this circle then contributes significantly to \textit{housing}, followed by the family circle contributing more to \textit{housing}. \textit{Tea art}-related articles shift between the family circle and the schoolmate circle. We also inspect other clusters and identify similar phenomenons, i.e. although different social circles tend to curate some relatively stable topics (e.g., traveling) with strong characteristics, some interests can shift between circles and the attention paid to them has its own ups and downs (noted as \textbf{F8}).

\par In the qualitative results for \textbf{RQ3}, 50.5\% of survey participants reported that leveraging different social circles as ``gatekeepers'' experienced some changes, with 32.4\% of them after they entered colleges, 36.1\% of them after they started to work, and 37.1\% when their lives shift to a new stage, such as getting married, becoming parents, and getting retired. The follow-up interviews complement some detailed explanations. ``\textit{now I prefer to read professional articles related to my major, and I will explore them by subscription accounts or my friend network,}'' reported by P1 (student, male); ``\textit{I read more articles related to my field so I pay more attention to my colleagues,}'' said P4 (male). ``\textit{After getting married, I consume more information about life, emotions, personal growth or the contents curated by my friends in similar circumstances. In fact, I have unsubscribed some subscription accounts about my original interests,}'' said P6 (recent married, female).

\section{Discussion}
\par The insights found in this study can provide implications for the future designs of socio-technical systems, e.g., social recommenders. Users tend to adjust their degree of reliance on subscription accounts or friend networks based on the quantity and quality of their content suppliers (\textbf{F1}). Meanwhile, elder users with fewer WeChat experience are more likely to consume friend-curated content (\textbf{F2}). For one thing, having more spare time, most of them enjoy socializing with old friends and classmates on social media (\textbf{F4}). As indicated in~\cite{madden2010older}, ``\textit{social media users are more likely to reconnect with people from their past, and these renewed connections provide a strong support network when people are near retirement.}'' For another, this group of users usually have strong information needs. Therefore, social recommender strategies for their contents, if applied, need to be tailored.

\par Regarding how gatekeeping gets reflected in different social circles and ties, survey participants indicate that the family circle cannot fully function as a gatekeeper, different from the quantitative analysis (\textbf{F3}). This is because of emotional supports, or because people may already have a pre-assumption of the contents curated by them. The qualitative study also finds that people's reading interests may shift over time due to (1) changes of their information needs and tastes may have altered; and (2) changes of their social circles composition around the same time, which is in accordance with \textbf{F7-8}: although users with similar friend composition tend to get stable curation for some information, some articles and the attention paid to them can shift between different social circles. The performance of DT regression indicates that the article consumption space preserves different social circles in a non-linear way, i.e., different social circles can share common information interests. People with a clear idea of what they want to read will go straight to appropriate gatekeepers known to fulfill the consumption demands. These gatekeepers can be either close friends with strong-ties or trustful followed subscription accounts. In addition, the quantitative analysis indicates that the information curation behaviors of friends with strong-ties hold implications for the unfollowed subscription outlet trust (\textbf{F5}). With a little help from these friends, people may be able to connect with new subscription accounts and improve their readiness to participate in an informed democracy. People are also willing to acquire new knowledge from weak-ties (\textbf{F6}), which bring serendipity. In either case, users manage different gatekeepers as content curators. We can, therefore, model pairwise friend influence and apply to potential recommendation scenarios such as the social advertisements to give more exposure to users who are more likely to consume friend-curated content.

\subsection{Limitation}
\par There are several limitations to this research. First, in \textbf{RQ3}, since we only consider the relationship between social circles and article categories, regression models may fail to capture other factors with a possibly high correlation between the social circle and article consumption space. As we only have a one-year longitudinal data due to high maintenance cost of our collaborators, and users are clustered only once based on their social composition, we cannot conclude that changes of social circles would influence information gatekeeping, since dramatic changes in users' social network or interests are unlikely to happen overnight. Second, in \textbf{RQ2}, we only use the number of common friends to measure tie strength. A more objective metric may be chatting frequency between two users. Third, in most cases, we just employ CTR to infer users' reading interests and do not dig into deeper the motivations behind their clicks by using other metrics.

\section{Conclusion}
\par In this paper, we conduct a mixed-methods approach to studying ``friend network as a latent gatekeeper'' phenomenon on a friend-based social media WeChat. We analyze over seven million users to infer how they accommodate and safeguard their information through social circles and social ties. We also conduct a survey of 216 WeChat users about their reading activities on WeChat. Results indicate that WeChat users prefer the friend network when information is overloaded. They like to leverage weak-ties getting exposed to new domains, and turn to strong-ties when demanding credible and reliable information. Elder users with fewer experience using WeChat are more likely to consume friend-curated contents. Users leverage social circles to gatekeep information interests and the interests and attention paid to them can shift from one social circle to another.

\begin{acks}
We are grateful for the valuable feedback and comments provided by the anonymous reviewers. This research was supported by WeChat-HKUST Joint Lab on AI Technology (WHATLAB) grant\#1617170-0 and HKUST-WeBank Joint Laboratory Project Grant No.: WEB19EG01-d.
\end{acks}

\bibliographystyle{ACM-Reference-Format}
\bibliography{sample-base}

\end{document}